\documentclass[letterpaper]{article} 
\usepackage{aaai2026}  
\usepackage{times}  
\usepackage{helvet}  
\usepackage{courier}  
\usepackage[hyphens]{url}  
\usepackage{graphicx} 
\usepackage{natbib}  
\usepackage{caption} 
\usepackage{booktabs}
\usepackage{multirow}
\usepackage{graphicx}
\usepackage{adjustbox}
\usepackage{amsmath} 
\usepackage{float}
\usepackage{algorithm}
\usepackage{algorithmic}
\usepackage{amssymb}
\usepackage{newfloat}
\usepackage{listings}
\DeclareCaptionStyle{ruled}{labelfont=normalfont,labelsep=colon,strut=off} 
\floatstyle{ruled}
\newfloat{listing}{tb}{lst}{}
\floatname{listing}{Listing}
\nocopyright

\title{USE: A Unified Model for Universal Sound Separation and Extraction}
\author{
    Hongyu Wang\textsuperscript{\rm 1,\rm 2}\equalcontrib,
    Chenda Li\textsuperscript{\rm 1,\rm 2}\equalcontrib \thanks{Corresponding authors},
    Xin Zhou\textsuperscript{\rm 1,\rm 2},
    Shuai Wang\textsuperscript{\rm 3,\rm 4},
    Yanmin Qian\textsuperscript{\rm 1,\rm 2}\footnotemark[\value{footnote}]
}
\affiliations{
    \textsuperscript{\rm 1}Auditory Cognition and Computational Acoustics Lab\\
MoE Key Lab of Artificial Intelligence, AI Institute\\
School of Computer Science, Shanghai Jiao Tong University, Shanghai, China\\

    \textsuperscript{\rm 2}VUI Labs, China \\
    \textsuperscript{\rm 3}Nanjing University, Suzhou, China \\
    \textsuperscript{\rm 4}Shenzhen Loop Area Institute, Shenzhen, China\\
    {\fontsize{10pt}{12pt}\selectfont
\{hongyu.wang, lichenda1996, zxzx818, yanminqian\}@sjtu.edu.cn\\
\{shuaiwang\}@nju.edu.cn}

}

\usepackage{bibentry}

\begin{document}

\maketitle

\begin{abstract}
Sound separation (SS) and target sound extraction (TSE) are fundamental techniques for addressing complex acoustic scenarios. While existing SS methods struggle with determining the unknown number of sound sources, TSE approaches require precisely specified clues to achieve optimal performance. This paper proposes a unified framework that synergistically combines SS and TSE to overcome their individual limitations. Our architecture employs two complementary components: 1) An Encoder-Decoder Attractor (EDA) network that automatically infers both the source count and corresponding acoustic clues for SS, and 2) A multi-modal fusion network that precisely interprets diverse user-provided clues (acoustic, semantic, or visual) for TSE. Through joint training with cross-task consistency constraints, we establish a unified latent space that bridges both paradigms. During inference, the system adaptively operates in either fully autonomous SS mode or clue-driven TSE mode. Experiments demonstrate remarkable performance in both tasks, with notable improvements of 1.4 dB SDR improvement in SS compared to baseline and 86\% TSE accuracy.

\end{abstract}
\begin{links}
    \link{Demo}{https://hongyuwang414.github.io/USE-demo/}
\end{links}
\section{Introduction}
In complex acoustic environments, Universal Sound Separation(USS) is capable of providing high-quality inputs for subsequent audio analysis tasks. USS has become a core task \cite{kavalerov2019universal,tzinis2020improving,tzinis2022compute,liu2024audio,pons2024gass, zhao2024universal,kong2023universal} and aims to isolate arbitrary types of sound sources, including speech, music, environmental sounds, and musical instruments, from complex audio mixtures without relying on predefined numbers or types of sound sources. This flexibility greatly expands the application scope of sound processing, making it indispensable in fields such as environmental monitoring and multimedia content analysis.

Despite recent advances in sound separation (SS) models, existing methods still face several limitations. For instance, many models require the number of sound sources in a mixture to be predefined during inference \cite{luo2019conv,zhao2024mossformer2,subakan2021attention}, which restricts their applicability in real-world scenarios where the number of sound sources is often uncertain. 
To address these issues, target sound extraction (TSE) utilizes some prior knowledge about the target sound from the mixture of an unknown number of sources \cite{liu2024separate, liu2024audio, zhao2018sound,liu2022separate, kilgour2022text, zhang2024multi}.
The prior knowledge about the target sound, also known as \textit{clue}.
For example, the AudioSep model \cite{liu2024separate} has shown remarkable performance by using the natural language description as the extraction clue; APT \cite{liu2024audio} uses audio samples as the clue for the TSE task; PixelPlayer \cite{zhao2018sound} uses visual information to extract target sounds.

Although target sound extraction can theoretically solve the problem of the unknown number of sources in the mixture, it suffers from two drawbacks in real applications: 1) First, the clues that are assumed to exist in advance may be of low quality or not be found in actual applications, which will degrade the performance of the TSE or even extract the wrong target.
2) The second drawback is that some previous studies \cite{delcroix2020improving,elminshawi2022new} 
found that the TSE-based methods usually achieve less performance than the SS methods when the number of sources is determined. 

\begin{figure}[t]
  \centering
  \includegraphics[height=0.65\linewidth,] {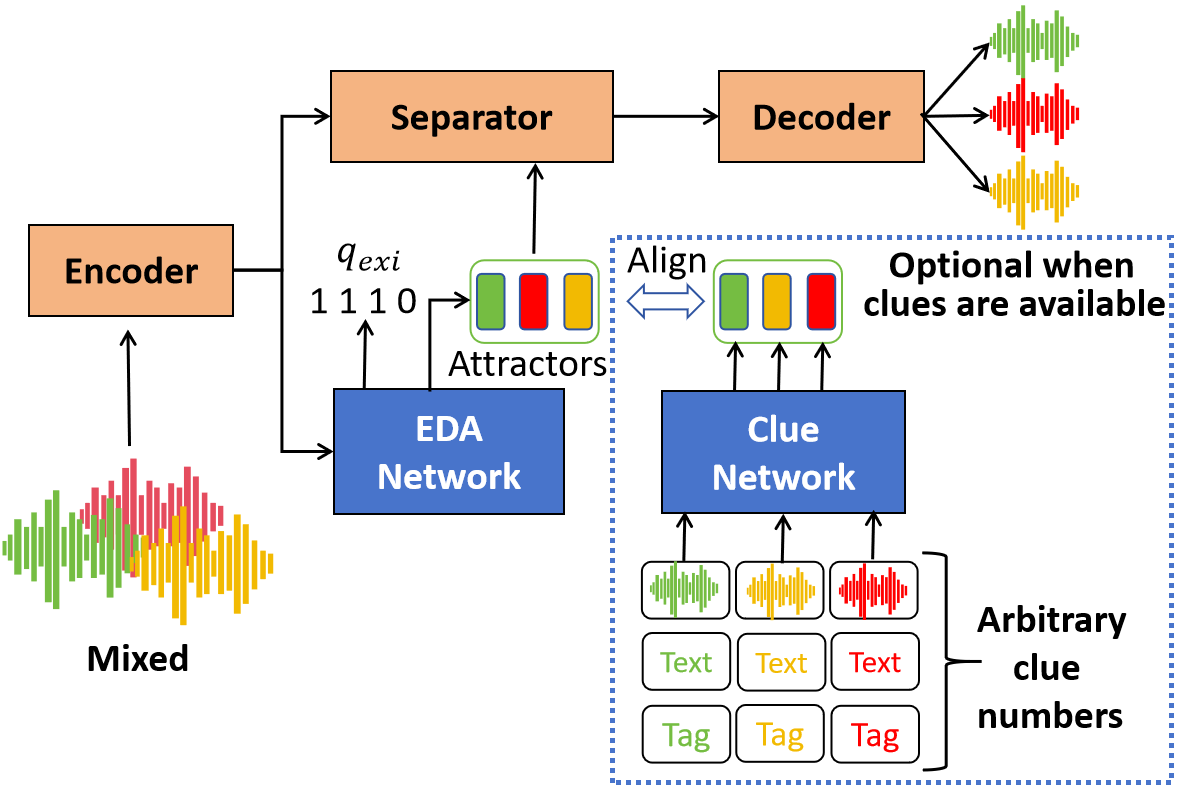} 
 \caption{Architecture of the USE Framework. Without clues, the EDA Network implicitly learns to represent all independent sound sources in the mixture as attractors, which are then used for separation. With arbitrary clues, the Clue Network generates Clue embeddings that are sent into the separation network for extraction. During training, semantic alignment between Attractors and Clue embeddings bridges the gap between Sound Separation (SS) and Target Sound Extraction (TSE).}
  \label{fig:The architecture of USE}
\end{figure}

In this paper, we propose Universal sound Separation and target sound Extraction (USE), a unified model, to solve the above-mentioned drawbacks of SS and TSE in complex acoustic scenarios. 
The USE comprises a separation backbone, an EDA module, and a multi-modal clue module. 
Without knowing the number of sound sources, the USE model can complete the SS task with a built-in EDA module.
Following the previous work \cite{li2023target}, the multi-modal clue module is robust for missing or low-quality clues.
It takes an uncertain number of clues of different modalities as input and generates a clue embedding in a unified space.
When there is any available modality of clues, the USE model can accomplish the TSE task with the clue module. 
In the training stage, we constrain the EDA embeddings and clue embeddings into a unified space to align them in a semantic space, which allows the USE model to perform SS or TSE tasks flexibly in the inference stage. We conducted the experiments on universal sound datasets with clues of multiple modalities, which shows the robustness of USE in complex acoustic scenarios. 

The main novelty and contributions of this paper are as follows: 1) We propose USE, a unified framework for SS and TSE, which can separate or extract sounds from a mixture of an unknown number of sources; 2) We show that the multi-task USE brings better performance on both SS and TSE baselines. 3) We conduct the experiments on a universal sound separation dataset with low-quality clues of multiple modalities, which shows the robustness of USE in complex acoustic scenarios.

\begin{figure*}[h]
  \centering
  \includegraphics[width=0.9\linewidth]{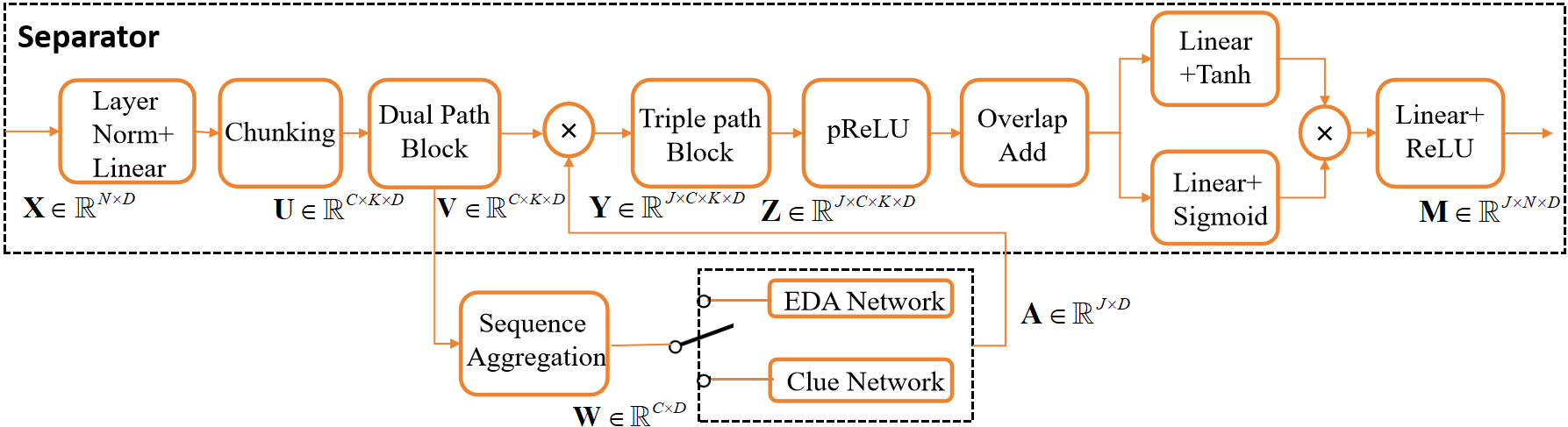}
  \caption{The architecture of the Sepformer-based Separator. It incorporates sequence aggregation to generate \( \mathbf{W} \). For the separation task, it feeds the EDA module to estimate attractors for each sound source in the mixture. For the TSE task, it integrates the Clue Network and features from other modalities for attention computation.}
  \label{fig:The architecture of the Separator}
\end{figure*}
\section{Related Work}
There are some attempts to solve the problems of SS and TSE mentioned above. For example, OR-PIT \cite{takahashi2019recursive} trains an independent binary classifier to determine whether speech should be further separated from the mixture;
In \cite{li2023target, cheng2024omnisep}, the impact of low-quality clues and clue missing are alleviated by utilizing the multi-modal clue; \cite{chetupalli2022speech,lee2024boosting} proposed to use an Encoder-Decoder-Attractor to count the number of sources in the SS task. While these approaches have effectively improved performance in single-task SS and TSE, unifying these tasks remains a challenge. Although attempts like those in \cite{saijo2023single, chetupalli2024unified} have used attractors to integrate SS and TSE, their methods are still relatively superficial. They fail to align attractors with clues in a semantic space, leading to a disjointed performance when handling both SS and TSE tasks, and they only focus on clean human speech rather than universal sound.

The methods proposed in \cite{saijo2023single} and \cite{chetupalli2024unified} struggle to accurately associate the target sound sources with the given clues during the TSE task.
In \cite{saijo2023single}, the TSE inference process involves weighting the attractors based on the weighted coefficients of the clues and attractors to extract the target audio. This method fails to effectively achieve semantic unification between attractors and clues. Additionally, it introduces extra computational overhead, particularly when dealing with a large number of sources in the mixture. Moreover, it relies solely on a single type of target source-related clue. When the quality of the clue is low or even absent, the TSE task cannot be effectively accomplished.
In \cite{chetupalli2024unified}, although the TSE task incorporates classification learning of audio source in attractors, it still does not consider the substitutability between attractors and clue embeddings. Moreover, it does not address the issue of low-quality or missing clues.

\section{Methods}
As shown in Figure~\ref{fig:The architecture of USE}, we propose a universal model for SS and TSE.
The model comprises a backbone network for sound separation, an Encoder-Decoder Based Attractors (EDA) network, and a multi-modal clue network. The EDA network estimates the number of sounds and their attractor embeddings from the mixed audio. 
The multi-modal clue network takes a variable number of clues with different modalities as input and estimates a unified clue embedding from the clues of different modalities.
Align loss is applied between the attractor embeddings and clue embeddings in the training.
In the inference stage, the separator can work with the EDA module to perform the SS task; if any clue about the target sound exists, the TSE task can be performed by replacing the attractor embedding with the clue embedding. 
We will elaborate on the separation network, EDA network, multi-modal clue network, training objectives, and training strategy in the following subsections.

\textbf{Sound Separation}. The sound separation network employs an Encoder-Separator-Decoder structure. Given a single-channel mixed signal \(\mathbf{x} \in \mathbb{R}^L\) (where \(L\) is of arbitrary length) from \(J\) sound sources \(\mathbf{S} = \{\mathbf{s}_j\}_{j=1}^{J} \), the network aims to estimate each source signal \(\hat{\mathbf{s}}_j\) based on \(\mathbf{x}\) and the estimated number of sources \(\hat{J}\) from the EDA network.

The encoder uses a 1-D convolution layer with ReLU activation to generate the hidden features \( \mathbf{X} \in \mathbb{R}^{N \times D}_+ \) of the mixed signal. The separator estimates a mask \( \mathbf{M} = \{\mathbf{m}_{j}\}_{j=1}^{\hat{J}} \) for each source. By element-wise multiplication of \( \mathbf{X} \) with \( \mathbf{m}_{j} \), we obtain the estimated hidden representations \( \hat{\mathbf{X}} = \{x_j\}_{j=1}^{\hat{J}} \). Finally, the decoder, composed of transposed convolution layers and symmetric to the encoder, estimates the separated sources \(\{\hat{\mathbf{s}}_j \in \mathbb{R}^L\}_{j=1}^{\hat{J}} \).

We employed both a time-domain model based on SepFormer \cite{subakan2021attention} and a frequency-domain model based on BSRNN \cite{luo2023music} as separators to validate the effectiveness of our joint training framework. Here, we primarily focus on the one based on SepFormer as shown in Figure~\ref{fig:The architecture of the Separator}. The input feature \( \mathbf{X} \) is first processed through a Layer-Norm layer and a linear layer, then segmented into overlapping chunks of size \( K = 250 \) with a 50\% overlap rate. The chunks are fed into a dual-path module, which integrates intra-chunk and inter-chunk transformer blocks~\cite{subakan2021attention,chetupalli2022speech}. Sequence aggregation leverages learnable weights to fuse information across distinct feature spaces and the output \( \mathbf{W} \) is fed into the EDA block or modality-variant clue network.

The dual-path block's output \( \mathbf{V} \) is element-wise multiplied with the source sound representations \( \mathbf{A} \in \mathbb{R}^{J \times D} \) from the EDA or Clue network, forming the input \( \mathbf{Y} \in \mathbb{R}^{J \times C \times K \times D} \) for the Triple Path Block. This block extends the dual-path design by adding an inter-channel transformer block, capturing relationships across channels. The final output \( \mathbf{Z} \in \mathbb{R}^{J \times C \times K \times D} \) is processed through a Parametric ReLU layer, followed by overlap-add (OVA) and a gated output layer with two linear layers. The final masks \( m_j \) are generated via a linear layer with ReLU activation.


\textbf{EDA Network}. The EDA network estimates the number and embeddings of distinct sound categories in mixed audio, enabling sound separation without prior knowledge of source counts. It employs an LSTM Encoder-Decoder framework \cite{hochreiter1997long} to convert frame-wise embeddings into global attractors.

The LSTM encoder updates its hidden state \( \mathbf{h}_t^{\text{enc}} \) and cell state \( \mathbf{c}_t^{\text{enc}} \) using the following equations:
\begin{equation}
\mathbf{h}_t^{\text{enc}}, \mathbf{c}_t^{\text{enc}} = h^{\text{enc}}(\mathbf{W}_t, \mathbf{h}_{t-1}^{\text{enc}}, \mathbf{c}_{t-1}^{\text{enc}}) \quad (t = 1, \ldots, C).
\end{equation}
The hidden and cell states of the encoder are initialized to zero vectors: \( \mathbf{h}_0^{\text{enc}} = \mathbf{0} \) and \( \mathbf{c}_0^{\text{enc}} = \mathbf{0} \).

The LSTM decoder \( \mathbf{h}_s^{\text{dec}} \) estimates global-wise attractors as:
\begin{equation}
\mathbf{h}_s^{\text{dec}}, \mathbf{c}_s^{\text{dec}} = h^{\text{dec}}(\mathbf{0}, \mathbf{h}_{s-1}^{\text{dec}}, \mathbf{c}_{s-1}^{\text{dec}}) \quad (s = 1, 2, \ldots).
\end{equation}
At each step, the hidden state \( \mathbf{h}_s^{\text{dec}} =: \mathbf{a}_s \in (-1, 1)^D \) serves as the attractor for sound category \( s \), with the dimensionality \( D \) matching that of the frame-wise embeddings \( \mathbf{W}_t \). The decoder's hidden and cell states are initialized by the encoder's final states: \( \mathbf{h}_0^{\text{dec}} = \mathbf{h}_T^{\text{enc}} \) and \( \mathbf{c}_0^{\text{dec}} = \mathbf{c}_T^{\text{enc}} \).

We compute the attractor existence probabilities using a fully connected layer with a sigmoid activation function, as shown in the following equation:
\begin{equation}
p_{\text{exi}} = \sigma \left( \mathbf{w}_{\text{exi}}^\top \mathbf{a}_s + b_{\text{exi}} \right),
\end{equation}
where \( \mathbf{w}_{\text{exi}} \in \mathbb{R}^D \) and \( b_{\text{exi}} \in \mathbb{R} \) are the trainable weights and bias parameters of the fully connected layer, respectively. We compare each \( p_{\text{exi}} \) with a predefined threshold \( \theta \) in the inference stage. If \( p_{\text{exi}} > \theta \), we consider the attractor to exist; otherwise, it is considered that no more sound sources exist in the mixture. The result of the existence check is denoted by \( q_{\text{exi}} \).

\begin{figure}[t]
  \centering
  \includegraphics[width=1\linewidth] {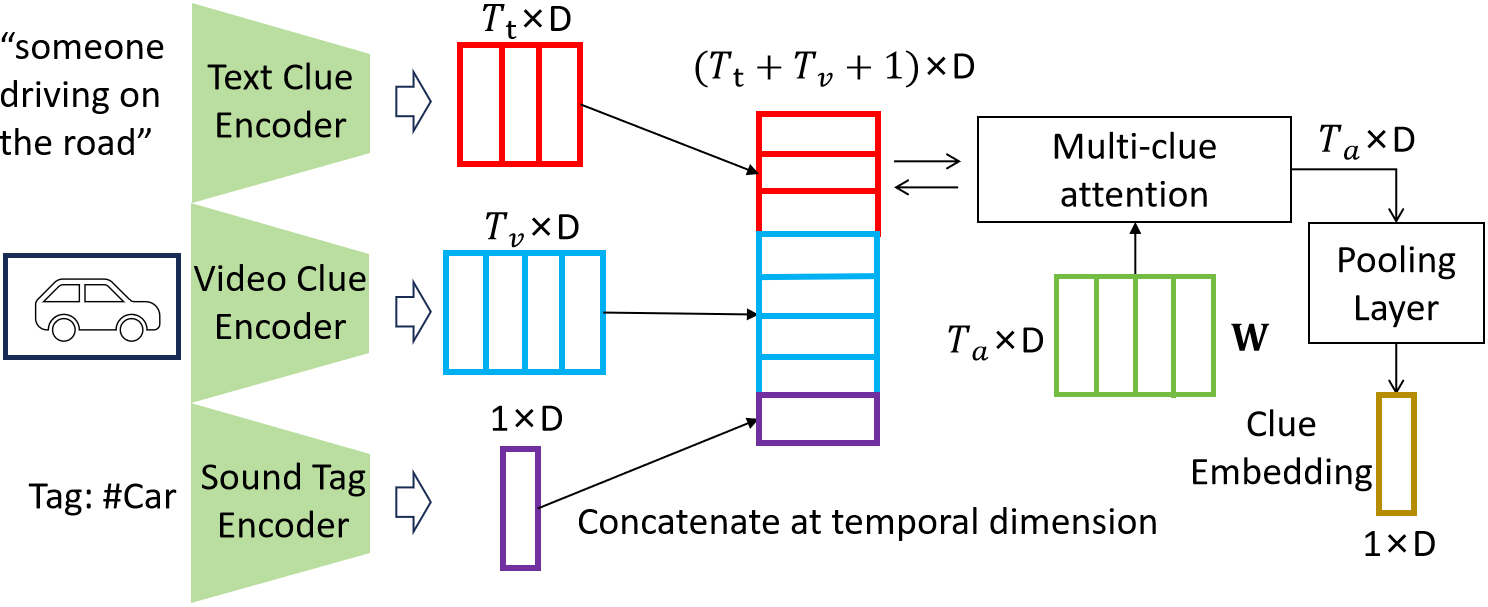} 
  \caption{Clue Network Architecture. When any one to three modalities of text, video, and sound tags are present, we employ dedicated pre-trained encoders to extract features from each modality. These features are then concatenated along the temporal dimension and, together with the output W from sequence aggregation, passed through multi-clue attention and pooling layers to generate the final Clue Embedding.}
  \label{fig:The architecture of the Clue network}
\end{figure}
\textbf{Clue Network}. As shown in Figure~\ref{fig:The architecture of the Clue network}, the clue network takes variable modality clues (e.g., text, video, sound) as input and generates a comprehensive clue embedding. Each modality is encoded into a unified \(D\) dimensional space using dedicated encoders. An attention module then integrates these clues into a temporally aligned fused clue, from which global clue information is derived by averaging over time.

The encoders convert various modalities into \(D\)-dimensional embeddings for target sound extraction. The text encoder uses a pre-trained DistilBERT model \cite{sanh2019distilbert} to transform natural language descriptions into embeddings \( \mathbf{O} \in \mathbb{R}^{T_t \times D} \), where \( T_t \) is the number of word tokens. The video encoder processes frames with a pre-trained Swin Transformer \cite{liu2021swin}, producing video embeddings \( \mathbf{V} \in \mathbb{R}^{T_v \times D} \), where \( T_v \) is the number of frames. The sound encoder maps one-hot encoded sound event tags to embedding vectors \( \mathbf{E} \in \mathbb{R}^{1 \times D} \). All the encoders have their parameters frozen during the training phase.

\textbf{Multi-Modal Concatenation:} The unified multi-modal clue \( \mathbf{U} \) is formed by concatenating the text, video, and sound tag embeddings:
\begin{equation}
\mathbf{U} = \text{Concatenate}(\mathbf{O}; \mathbf{V}; \mathbf{E}) \in \mathbb{R}^{(T_t + T_v + 1) \times D}.
\end{equation}

\textbf{Attention-Based Clue Fusion:} Utilizing the output of Sequence Aggregation \( \mathbf{W} \) as the query and \( \mathbf{U} \) as both the key and value in the attention mechanism, we derive the fused clue \( \mathbf{C}_u \) as follows:
\begin{equation}
\mathbf{C}_u = \text{MultiHeadAttention}(\mathbf{W}, \mathbf{U}, \mathbf{U}) \in \mathbb{R}^{T_a \times D},
\end{equation}
where \( \text{MultiHeadAttention}(\mathbf{Q}, \mathbf{K}, \mathbf{V}) \) denotes a multi-head attention mechanism \cite{li2023target, waswani2017attention} with query \( \mathbf{Q} \), key \( \mathbf{K} \), and value \( \mathbf{V} \). It is evident that the fused clue \( \mathbf{C}_u \) shares the same length \( T_a \) as the sound embedding \( \mathbf{Q} \).

\subsection{Loss Function}
The objective function is a combination of three distinct losses: the source separation loss \( \mathcal{L}_{\text{sep}} \), the source counting loss \( \mathcal{L}_{\text{count}} \), and  the align loss \( \mathcal{L}_{\text{align}} \) between attractors and clues. Specifically, the alignment loss \( \mathcal{L}_{\text{align}} \) is defined as:
\begin{equation}
\mathcal{L}_{\text{align}} = \mathcal{L}_{\text{MSE}} + \mathcal{L}_{\text{InfoNCE}},
\end{equation}
where \( \mathcal{L}_{\text{MSE}} \) denotes the Mean Squared Error loss and \( \mathcal{L}_{\text{InfoNCE}} \) denotes the InfoNCE loss. Our preliminary experiments showed that using only the MSE Loss lowers attractor–clue alignment accuracy and degrades separation metric SI-SNRi, whereas relying solely on the InfoNCE loss results in slow convergence.


\textit{Source Separation Loss (\( \mathcal{L}_{\text{sep}} \))}: In multi-source separation tasks, we apply Signal-to-Noise Ratio (SNR) with Permutation Invariant Training (PIT) as the objective function:
\begin{equation}
\mathcal{L}_{\text{sep}} = -\max_{\pi \in \Pi_K} \sum_{k=1}^{K} 10 \log_{10} \left( \frac{\sum_{t=1}^{T} \mathbf{s}_k(t)^2}{\sum_{t=1}^{T} (\hat{\mathbf{s}}_{\pi(k)}(t) - \mathbf{s}_k(t))^2} \right), 
\end{equation}
where \( \Pi_K \) represents all possible permutations, \( \pi \) is a permutation mapping, \( K \) is the number of target sources, \( T \) is the length of the signal, \( \hat{\mathbf{s}}_{\pi(k)}(t)\) is the value of the estimated source at time step \( t \) under permutation \( \pi \), and \( \mathbf{s}_k(t) \) is the value of the target source at time step \( t \).

\textit{Source Counting Loss (\( \mathcal{L}_{\text{count}} \))}: This loss is determined using the binary cross-entropy measure, which evaluates the accuracy of the estimated number of sound sources in mixture, given by:
\begin{equation}
\mathcal{L}_{\text{count}} = - \sum_{i=1}^{K} y_i \log(p_{\text{exi}}) + (1 - y_i) \log(1 - p_{\text{exi}})
\end{equation}
where \( y_i \) is the true label and $p_{\text{exi}}$ is the predicted probability.

\textit{MSE Loss (\( \mathcal{L}_{\text{MSE}} \))}: The MSE Loss \( \mathcal{L}_{\text{MSE}} \) is calculated by first identifying the optimal permutation between the attractors and clues according to final Separation PIT loss. Let \( \pi^{*} \) denote the best permutation in the Equation (7) and \( D \) be the dimension of the embeddings. \( \mathbf{a}_{\pi^{*}(m),i} \) represents the \( i \)-th value of the \( m \)-th attractor under permutation \( \pi^{*} \), while \( \mathbf{c}_{m,i} \) is the \( i \)-th value of the \( m \)-th clue. \( M \) is the total number of clues or attractors. The MSE Loss can be expressed as:
\begin{equation}
\mathcal{L}_{\text{MSE}} = \frac{1}{M} \sum_{m=1}^{M} \frac{1}{D} \sum_{i=1}^{D} \left( \mathbf{a}_{\pi^{*}(m),i} - \mathbf{c}_{m,i} \right)^2
\end{equation}

\textit{InfoNCE Loss (\( \mathcal{L}_{\text{InfoNCE}} \))}: Given \(N\) attractors \( \{\mathbf{a}_i\}_{i=1}^{N} \) and \(N\) clues \( \{\mathbf{c}_j\}_{j=1}^{N} \), and the best permutation \( \pi^{*} \) obtained from the Equation (7), the InfoNCE Loss \( \mathcal{L}_{\text{InfoNCE}} \) is computed as the average of the InfoNCE loss \cite{oord2018representation} for each corresponding pair:
\begin{equation}
\mathcal{L}_{\text{InfoNCE}} = - \frac{1}{N} \sum_{i=1}^{N} \log \frac{\exp(\mathbf{z}_{a_i} \cdot \mathbf{z}_{c_{\pi^{*}(i)}} / \tau)}{\sum_{j=1}^{N} \exp(\mathbf{z}_{a_i} \cdot \mathbf{z}_{c_j} / \tau)}
\end{equation}

where \( \mathbf{z}_{a_i} \) is the embedding of the \(i\)-th attractor. \( \mathbf{z}_{c_j} \) is the embedding of the \(j\)-th clue. \( \pi^{*}(i) \) is the index of the clue that is optimally matched to the \(i\)-th attractor according to the Separation PIT loss. \( \tau \) is the temperature parameter.

\subsection{Training and Inference Strategies}
We applied a two-stage training method for model optimization. In stage 1, we train the model for sound separation with an EDA network to estimate the number of sources and compute the source counting loss \(\mathcal{L}_{\text{count}}\). Then, the EDA representations are fed into the separation network, which generates separated sounds for calculating the separation loss \(\mathcal{L}_{\text{sep}}\). This trains the separation and EDA networks jointly.

In stage 2, we randomly select EDA representations (30\% chance) or TSE-generated clue embeddings (70\% chance) as inputs to the separation network. For TSE, we train with all seven combinations of modalities (present or absent) equally, using the \(\mathcal{L}_{\text{sep}}\) loss in a fixed order. We align the two embedding to learn via align loss \(\mathcal{L}_{\text{align}}\). This strategy enhances adaptability and robustness.

\textbf{Inference Phase:} In the absence of clues, the EDA module estimates the number and representation of all independent sound sources in the mixture as attractors, which are then fed into the Separator for separation. When one to three modalities of clues are available, the Clue Network generates Clue Embeddings based on these clues, which are subsequently sent into the Separator for targeted extraction.

\section{Experiments}
\subsection{Datasets}
To compare our proposed model with existing models, we used the same universal sound dataset as in the \cite{li2023target} paper, which is based on AudioSet \cite{gemmeke2017audio}. AudioSet is a large-scale dataset extracted from YouTube, containing 527 sound classes with weak labels. Each 10-second clip typically includes multiple sound events without precise timing annotations. To isolate single sound sources, we followed the preprocessing method in \cite{kong2020source}, using a pre-trained Sound Event Detection (SED) model \cite{kong2020panns} to identify sound event anchors and extract 2-second audio segments.

We also expanded the 2Mix dataset from \cite{li2023target} to create a 2\&3Mix dataset by remixing existing audio sources, generating 248k mixed samples (140 hours) with 2 or 3 sound sources for training. In the subsequent sections, we will refer to the test sets as Seen datasets. For both the validation and Seen sets, we created 2Mix and 3Mix scenarios. Specifically, the validation sets included 0.5 hours of data, while the Seen datasets consisted of 1 hour of data for each scenario. Additionally, we generated 0.7 hours of data with unseen sound classes (mostly musical instruments) for each scenario to assess the model.

For text clues, we used an audio captioning model \cite{wu2019audio} to generate pseudo-natural descriptions from 2-second audio clips. For visual clues, we extracted frames from the aligned 2-second video segments (15 FPS). For tag clues, we converted SED probabilities into one-hot vectors.

\subsection{Training and Evaluation Details}
The experiments were conducted using the ESPnet-SE toolkit \cite{li2021espnet}. 
The training was divided into two stages: In stage 1, the learning rate was set to \(10^{-4}\) for 70 epochs, and in stage 2, it was adjusted to \(3 \times 10^{-5}\) for an additional 30 epochs. The threshold $\theta$ of EDA is set to 0.5 and the clue embedding dimension D is set to 256.

We used the SNR improvement (SNRi) to evaluate source separation performance. When the estimated number of sources mismatched the ground truth, we adopted the following approach: If the number of sources was over-estimated, we retained only the first k estimates for evaluation, where k equals the ground-truth number of sources. If under-estimated, we used a silence signal (all zeros) as the estimate for the missing sources.
\subsection{Separation Results}
\begin{table}[th]
  \centering
  \begin{tabular}{lccc}
    \toprule
      \textbf{Model} & \textbf{Parameters(M)} & \textbf{2 Mix} & \textbf{3 Mix}  \\
    \midrule
     TDANet-Wav & 10.8& 17.5 & / \\
     TDANet-STFT & 7.4& 12.7 & / \\
     BSRNN-Large & 21.8& 15.2 & / \\
     USE-B& 8.1& \textbf{17.8}  &15.0 \\
     USE-B$^{*}$  & 8.1& 17.7  &15.0 \\
    \bottomrule
  \end{tabular}
   \caption{Speech separation results on Libri2Mix and our simulated  Libri3Mix (SI-SNRi / dB) for USE-B versus competitive models. $^{*}$ represents separation under an unknown and variable number of speakers.}
  \label{tab:Libri2Mix}
\end{table}

\begin{table}[th]
  \centering
  \begin{tabular}{lcccc}
    \toprule
    \textbf{System} & \multicolumn{2}{c}{\textbf{2 Mix}} & \multicolumn{2}{c}{\textbf{3 Mix}} \\
    \cmidrule(lr){2-3} \cmidrule(lr){4-5}
    & \textbf{Seen} & \textbf{Unseen} & \textbf{Seen} & \textbf{Unseen} \\
    \midrule
    Sepformer  & 7.4 & 6.6 & / & / \\
    USE-S(stage 1) & 8.7 & 7.9 & 6.4 & 5.2 \\
    USE-S(stage 2) & \textbf{8.8} & \textbf{8.2} & \textbf{7.2} & \textbf{6.3} \\
     \midrule
    BSRNN  & 8.6 & 8.4 & / & / \\
    USE-B(stage 1) & 9.2 & 8.6 & 5.9 & 4.8 \\
    USE-B(stage 2) &\textbf{9.4}  & \textbf{8.8}&\textbf{6.3}  &\textbf{4.8} \\
    \bottomrule
  \end{tabular}
  \caption{Universal sound separation results for USE-S, USE-B and baseline models (SNRi/dB) on AudioSet dataset, together with ablation studies on EDA Network (stage 1) and Multi-task training strategy (stage 2).}
  \label{tab:separation_results}
\end{table}

\begin{table*}[htbp]
\centering
\begin{tabular}{lcc|cc|cc|cc}
    \toprule
    \multicolumn{3}{c}{{Available Clues}} & \multicolumn{2}{c}{{DCCRN}\cite{li2023target}} & \multicolumn{2}{c}{{USE-S}} & 
    \multicolumn{2}{c}{{USE-B}}\\
    \cmidrule(lr){1-3} \cmidrule(lr){4-5} \cmidrule(lr){6-7} \cmidrule(lr){8-9}
    tag & text & video & Seen & Unseen & Seen & Unseen & Seen & Unseen\\
    \midrule
    \checkmark & \checkmark & \checkmark & 6.9 & 6.5 &8.5 &8.1&\textbf{8.9} &\textbf{8.8}\\
    \checkmark & \checkmark & & 6.8 & 6.4  &8.5 &7.9&8.6&8.7\\
     &\checkmark & \checkmark & 6.5 & 6.4  &8.2 &8.0&8.6&8.7\\
     \checkmark& & \checkmark& 6.6 & 6.4  &7.8 &7.8&8.2&8.4\\
    \checkmark &  & & 6.4 & 6.2  &7.3 &7.2&7.4&8.0\\
    & \checkmark & & 6.3 & 6.0  & 8.2&7.8&8.0 &8.4\\
    & & \checkmark & 5.8 & 5.9  &6.8 &7.2&6.2 &7.4\\
    \bottomrule
\end{tabular}
\caption{TSE Performance comparison of different models with various weakly labelled clues (SNRi/dB).}
\label{tab:performance_comparison}
\end{table*}
\noindent \textbf{Speech Separation.} To assess the ability of the USE framework to perform separation under unknown and variable numbers of speakers, we first compare our unified framework USE-B  (based on BSRNN \cite{luo2023music}) against the leading universal sound separation models in \cite{pons2024gass} following the similar evaluation adopted in the work on the LibriMix dataset \cite{cosentino2020librimix}, as shown in Table~\ref{tab:Libri2Mix}. The results demonstrate that USE-B achieves superior separation performance with a minimal parameter count while remaining robust to both known and unknown numbers of speakers.


\noindent \textbf{Universal Sound Separation.} 
In the universal sound separation experiment, we employed baseline model \cite{subakan2021attention, luo2023music} trained on the 2Mix dataset and the separation results are as illustrated in the Table~\ref{tab:separation_results}. We also assessed the separation performance of models trained using our unified model USE-S (based on Sepformer \cite{subakan2021attention}) and USE-B (based on BSRNN \cite{luo2023music}). Our proposed model architecture differs in that it is not restricted to processing a fixed number of audio mixtures. So we trained our USE model on both 2Mix and 3Mix datasets (referred to as stage 1 training) and achieved better SNRi results, representing an improvement over baseline model and thereby validating the superiority of our strategy.


\begin{table}[th]
  \centering
  \begin{tabular}{lcccc}
    \toprule
    \textbf{System} & \multicolumn{2}{c}{\textbf{2 Mix}} & \multicolumn{2}{c}{\textbf{3 Mix}} \\
    \cmidrule(lr){2-3} \cmidrule(lr){4-5}
    & \textbf{Seen} & \textbf{Unseen} & \textbf{Seen} & \textbf{Unseen} \\
    \midrule
    Sepformer  &8.5  &8.0  &6.5  & 5.7 \\
    USE-S &8.5 &8.1  &6.5 &5.9  \\
    \midrule
    BSRNN &8.4 & 8.4 &5.9  &4.8 \\
    USE-B &8.9  & 8.8 &6.3  & 5.0 \\ 
    \bottomrule
  \end{tabular}
 \caption{TSE results for USE-S, USE-B and baseline models (SNRi/dB), together with ablation studies on Multi-task training strategy.}
  \label{tab:TSE_results}
\end{table}

In stage 2 of the training, we integrated TSE with the separation task, simultaneously training for both tasks. We evaluated the separation outcomes of stage 2, and the results showed a slight increase in SNRi. Additionally, the model in stage 2 is also capable of effectively performing TSE task. These findings further substantiate the rationality and effectiveness of our proposed unified training approach for TSE and Sound Separation (SS) tasks.


\subsection{Evaluation of Source Counting}
In our study, we evaluated the accuracy of source count estimation during attractor-based separation. During the inference phase, with the EDA threshold $\theta$ set to 0.5, the accuracy of source count estimation can reach over 80\% in 2Mix scenario and over 70\% in the 3-mix scenario, even on the highly noisy AudioSet dataset. This enables the USE framework to automatically assess and separate sound sources in complex scenarios with unknown numbers of sound sources.


\subsection{TSE Results}


\begin{table}[ht]
    \centering
    \begin{tabular}{lcc}
        \toprule
        \textbf{Model} & \textbf{2Mix} & \textbf{3Mix} \\
        \midrule
        MAE (audio) & $5.6$ & / \\
        USS (audio) & $5.6$ & / \\
        LASS (text) & $6.8$ & / \\
        Audiosep (text) & $7.7$ & / \\
        DCCRN (text+tag+video) & $6.9$ & / \\
        Sepformer (text+tag+video)  & $8.5$ & $6.5$ \\
        USE-S (text+tag+video)& $8.5$ & \textbf{6.5}\\
        BSRNN (text+tag+video) &8.4  &5.9 \\
        USE-B (text+tag+video)& \textbf{8.9} &$6.3$ \\
        \bottomrule
    \end{tabular}
 \caption{TSE Performance comparison  (SNRi/dB).}
    \label{tab:TSE_performance}
\end{table}

As shown in Table~\ref{tab:TSE_results}, we compared the extraction performance of models trained using our unified framework-—such as USE-S and USE-B--against baseline models trained solely for the TSE (Target Sound Extraction) task. The results show that after joint training with SS and TSE, the extraction performance of USE-S and USE-B did not decline. In fact, there was a slight improvement in TSE performance. Additionally, these models retained the ability to perform separation tasks using the EDA module in the absence of clues.

As shown in Table~\ref{tab:performance_comparison}, we compared USE-S and USE-B with DCCRN \cite{li2023target} using 2Mix data. Our model significantly outperforms DCCRN across all modality configurations, achieving the best performance when all three modalities are used together. The metric of USE-B reaches 8.9 and 8.8 on the Seen and Unseen dataset, respectively. This represents an improvement of 29.0\% and 35.4\% over DCCRN, respectively. Furthermore, our model also demonstrates strong performance under conditions where only single modality clue is available, thereby proving its superior robustness and adaptability.

As shown in Table~\ref{tab:TSE_performance}, we also compared the performance of different universal sound extraction models on the TSE task. Most of these models used the same SED cropping strategy \cite{kong2020source} on the AudioSet dataset as our processing method \cite{zhao2024universal, kong2023universal, liu2022separate, liu2024separate, li2023target, subakan2021attention, luo2023music}. However, there might be differences in the details , so these results should only be used as a reference. Nevertheless, it is evident from the table that our proposed USE-B model performs the best in TSE task, with performance comparable to that of the separately trained BSRNN+ClueNet model for only TSE task.

\begin{table*}[th]
  \centering
  \begin{tabular}{lcccccccccc}
    \toprule
    \textbf{Datasets} & \multicolumn{2}{c}{\textbf{2 Mix}} & \multicolumn{2}{c}{\textbf{3 Mix}} & \multicolumn{2}{c}{\textbf{4 Mix}} & \multicolumn{2}{c}{\textbf{5 Mix}} & \multicolumn{2}{c}{\textbf{6 Mix}}\\
    \cmidrule(lr){2-3} \cmidrule(lr){4-5} \cmidrule(lr){6-7} \cmidrule(lr){8-9} \cmidrule(lr){10-11}
    & \textbf{FN} & \textbf{PN} & \textbf{FN} & \textbf{PN} & \textbf{FN} & \textbf{PN} & \textbf{FN} & \textbf{PN} & \textbf{FN} & \textbf{PN}\\
    \midrule
     Universal-dataset &11.1 &11.7  &11.3  &11.4  & 10.4&10.3  &9.5  &9.4&9.0  &9.3\\
     Speech+Music &12.3 &12.3  &9.7  &9.3  &8.5 &8.2  &7.1  &7.0 &/  &/\\
    FUSS &14.4 &14.7  & 14.8 & 15.0 & 13.1&12.6  &/  &/ &/  &/\\
    Audioset(Seen)&8.9 &9.2  &6.7  & 6.3  &/ & / &/  &/ &/  &/\\
    Audioset(Unseen)&8.5 & 8.8 &5.9  & 5.3 &/ & / &/  &/ &/  &/\\
    \bottomrule
  \end{tabular}
\caption{Universal Sound Separation Performance of USE-B trained on multi-datasets (SNRi/dB), FN represents Fixed Number of sources, PN represents Predicted Number of sources.}
  \label{tab:6mixseparation_results}
\end{table*}

\subsection{Visualization of Attractors and Clues}
\begin{figure}[t]
  \centering
  \includegraphics[height=0.61\linewidth]{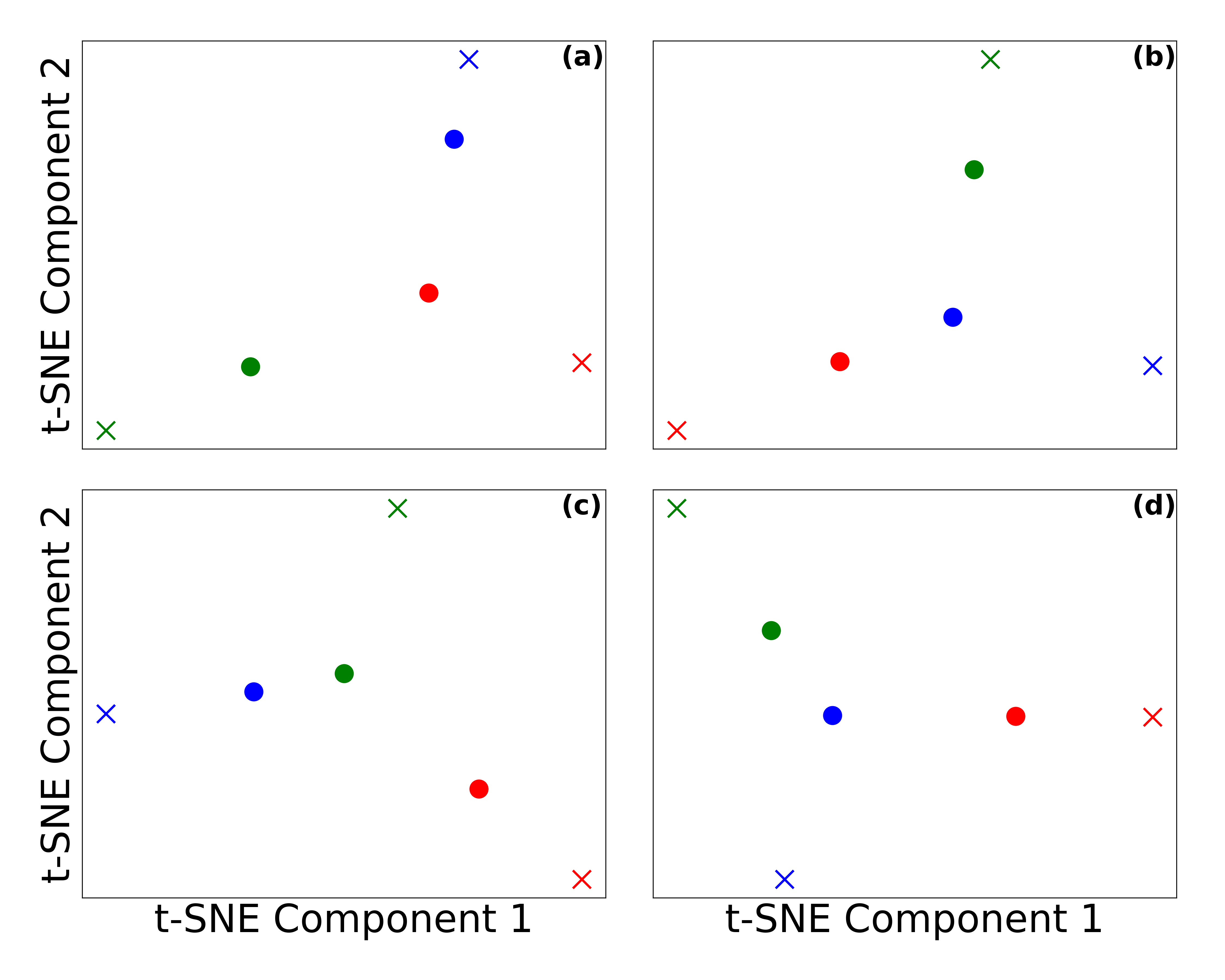}
  \caption{t-SNE visualization of attractors and clues. Different colors denote different sound-source types;
    circles ({$\bigcirc$}) represent attractors,
    crosses ({$\times$}) represent clues. (a) and (b) show the t-SNE visualizations of attractors and clues in the Audioset (Seen) 3-mix, while (c) and (d) show those in the Audioset (Unseen) 3-mix.}
  \label{fig:t-SNE Visualization of Attractors and Clues}
\end{figure}

During the inference process, we evaluated the accuracy of matching clues and attractors by classifying each attractor into one of the clue embeddings via the InfoNCE loss in Equation (10). The accuracy reaches 86.0\% in the 2-mix scenario and 65.3\% in the 3-mix scenario. This demonstrates that we have effectively unified attractors and clues in the semantic space, enabling them to substitute for each other in the absence of one or the other. This highlights the strong superiority and flexibility of our approach.

In addition, during the inference process, we randomly selected four 3Mix mixtures, two from the Seen dataset and two from the Unseen dataset. We performed t-SNE visualization \cite{van2008visualizing} for both attractors and clues in each mixture. As shown in the Figure~\ref{fig:t-SNE Visualization of Attractors and Clues}, different sound types have a certain spatial distance in the t-SNE plot, indicating that they possess a certain level of separability in the feature space. In addition, attractors and clues of the same sound type exhibit similarity within the feature space, indicating that they could serve as viable replacements for each other if one is absent.

\subsection{Universal Sound Separation}
We have additionally trained USE-B based on BSRNN on a large-scale general audio dataset, which includes the universal-dataset (2$\sim$6 mix) mainly remixed by VGGSound \cite{chen2020vggsound}, AudioSet \cite{gemmeke2017audio} (2$\sim$3 mix), FUSS \cite{wisdom2021s} (2$\sim$4 mix), and Musan \cite{snyder2015musan} (music) + Librispeech \cite{panayotov2015librispeech} (speech) (2$\sim$5 mix). During the evaluation, we considered two situations: one where the number of sources to be separated is known in advance, and another where the number of sources is unknown and needs to be estimated using the EDA module. The evaluation results are as Table~\ref{tab:6mixseparation_results}. Firstly, we can see that USE-B is capable of adapting to mixtures with varying numbers of sources and performs well across different datasets. Secondly, we have found that the method of predicting the number of sound sources using the EDA module generally matches or outperforms the approach based on a pre-determined fixed number of sources. 

We also compared our USE-B model with other competitive models on the FUSS \cite{wisdom2021s,wang2021sequential}(Dry, without reverberation) test set, as shown in Table~\ref{tab:FUSS_results}. It can be observed that our model achieved superior separation performance in both 2-mix, 3-mix, and especially in 4-mix scenarios.
\begin{figure}[h]
  \centering
  \includegraphics[height=0.57\linewidth]{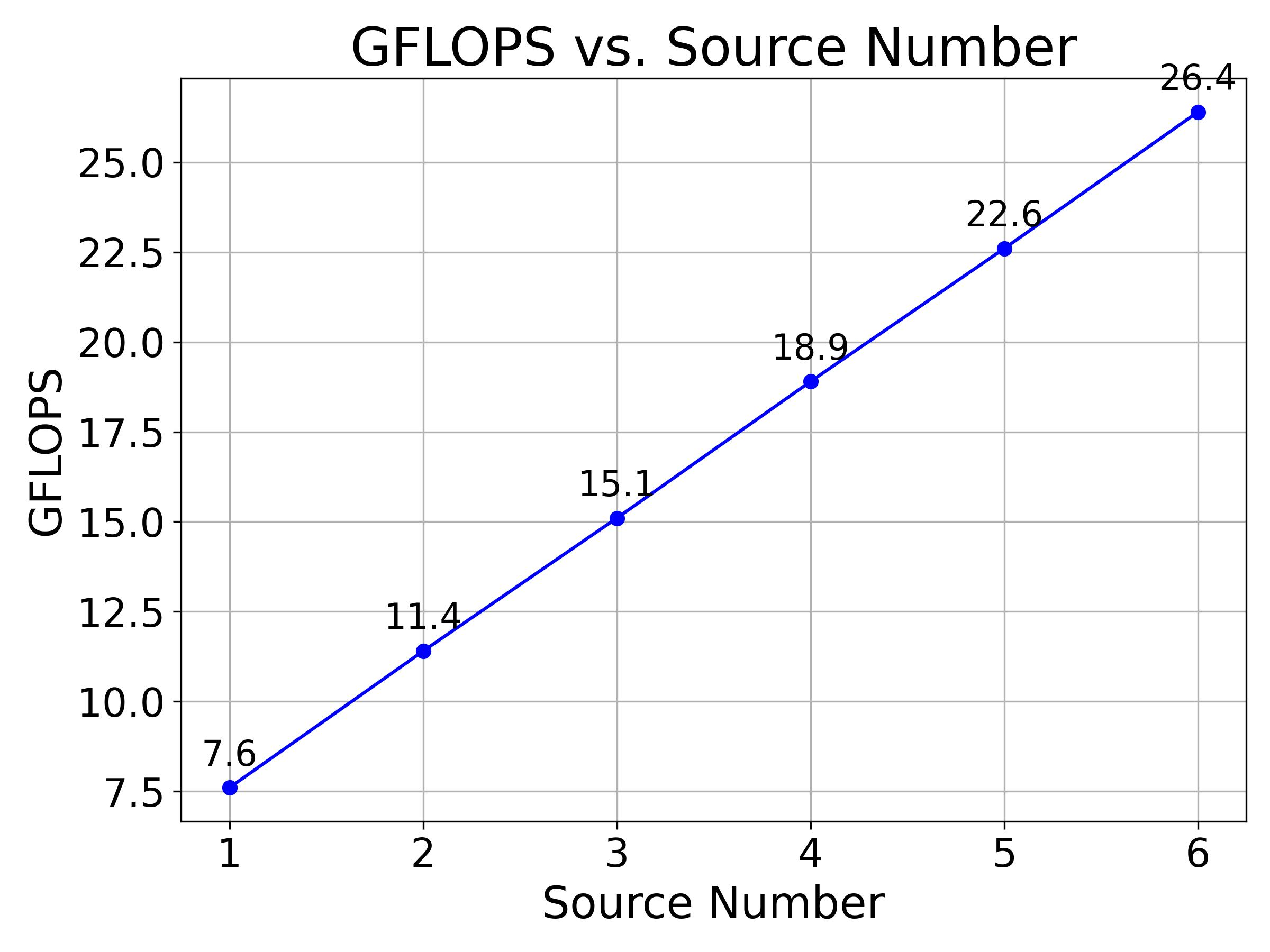}
  \caption{The GFLOPS of the USE Model as the number of sources increases.}
  \label{fig:GFLOPS}
\end{figure}

We have also evaluated the inference speed of extracting 1 to 6 source audio signals from a mixture as shown in Figure \ref{fig:GFLOPS}. It is observed that the computational complexity during inference increases linearly with the number of sound sources. Notably, even when inferring up to six sources, the computational demand in terms of GFLOPS remains below 30. This characteristic ensures the real-time performance and high efficiency of the USE-B model during inference.

\begin{table}[H]
  \centering
  \begin{tabular}{lccc}
    \toprule
      \textbf{Model} & \textbf{2 Mix} & \textbf{3 Mix} & \textbf{4 Mix}  \\
    \midrule
     TDCN++ & 11.2& 11.6 & 7.4\\
     USE-B& 12.8& 13.1  & 11.9\\
    \bottomrule
  \end{tabular}
 \caption{Universal Sound Separation Performance Comparison on FUSS (Dry) dataset (SI-SNRi/dB).}
  \label{tab:FUSS_results}
\end{table}

\section{Conclusion}
In this study, we introduced USE, a universal method that unifies the Sound Separation and Target Sound Extraction tasks. This novel approach can handle diverse sound types, variable numbers of sources, and multiple modalities of clues and outperforms existing models in multi-source separation and target sound extraction tasks thanks to its ability to integrate an arbitrary number of clues.
However, we found that the performance of USE can be influenced by the cleanliness of individual sound sources in the training data. One promising direction is to develop methods for dynamically adjusting the granularity of attractors to improve its adaptability to various tasks. In different acoustic scenarios, some sound sources need to be separated individually, while others should be grouped together. Additionally, integrating USE with sound event detection and understanding tasks could provide a more comprehensive approach to audio processing. By combining these capabilities, we can develop a more robust framework and offer deeper insights into the acoustic environment.

\section*{Acknowledgements}
This work was supported in part by China STI 2030-Major Projects under Grant No. 2021ZD0201500, and in part by China NSFC project under Grants No. U25A20409.
\bibliography{aaai2026}

%


\end{document}